\documentclass[aps,prl,reprint,superscriptaddress]{revtex4-1}
\usepackage{graphicx}  
\usepackage{subfigure}

\begin{document}

\title{Limits on Neutrino Emission from Gamma-Ray Bursts with the 40 String IceCube Detector}

\affiliation{III. Physikalisches Institut, RWTH Aachen University, D-52056 Aachen, Germany}
\affiliation{Dept.~of Physics and Astronomy, University of Alabama, Tuscaloosa, AL 35487, USA}
\affiliation{Dept.~of Physics and Astronomy, University of Alaska Anchorage, 3211 Providence Dr., Anchorage, AK 99508, USA}
\affiliation{CTSPS, Clark-Atlanta University, Atlanta, GA 30314, USA}
\affiliation{School of Physics and Center for Relativistic Astrophysics, Georgia Institute of Technology, Atlanta, GA 30332, USA}
\affiliation{Dept.~of Physics, Southern University, Baton Rouge, LA 70813, USA}
\affiliation{Dept.~of Physics, University of California, Berkeley, CA 94720, USA}
\affiliation{Lawrence Berkeley National Laboratory, Berkeley, CA 94720, USA}
\affiliation{Institut f\"ur Physik, Humboldt-Universit\"at zu Berlin, D-12489 Berlin, Germany}
\affiliation{Fakult\"at f\"ur Physik \& Astronomie, Ruhr-Universit\"at Bochum, D-44780 Bochum, Germany}
\affiliation{Physikalisches Institut, Universit\"at Bonn, Nussallee 12, D-53115 Bonn, Germany}
\affiliation{Dept.~of Physics, University of the West Indies, Cave Hill Campus, Bridgetown BB11000, Barbados}
\affiliation{Universit\'e Libre de Bruxelles, Science Faculty CP230, B-1050 Brussels, Belgium}
\affiliation{Vrije Universiteit Brussel, Dienst ELEM, B-1050 Brussels, Belgium}
\affiliation{Dept.~of Physics, Chiba University, Chiba 263-8522, Japan}
\affiliation{Dept.~of Physics and Astronomy, University of Canterbury, Private Bag 4800, Christchurch, New Zealand}
\affiliation{Dept.~of Physics, University of Maryland, College Park, MD 20742, USA}
\affiliation{Dept.~of Physics and Center for Cosmology and Astro-Particle Physics, Ohio State University, Columbus, OH 43210, USA}
\affiliation{Dept.~of Astronomy, Ohio State University, Columbus, OH 43210, USA}
\affiliation{Dept.~of Physics, TU Dortmund University, D-44221 Dortmund, Germany}
\affiliation{Dept.~of Physics, University of Alberta, Edmonton, Alberta, Canada T6G 2G7}
\affiliation{Dept.~of Subatomic and Radiation Physics, University of Gent, B-9000 Gent, Belgium}
\affiliation{Max-Planck-Institut f\"ur Kernphysik, D-69177 Heidelberg, Germany}
\affiliation{Dept.~of Physics and Astronomy, University of California, Irvine, CA 92697, USA}
\affiliation{Laboratory for High Energy Physics, \'Ecole Polytechnique F\'ed\'erale, CH-1015 Lausanne, Switzerland}
\affiliation{Dept.~of Physics and Astronomy, University of Kansas, Lawrence, KS 66045, USA}
\affiliation{Dept.~of Astronomy, University of Wisconsin, Madison, WI 53706, USA}
\affiliation{Dept.~of Physics, University of Wisconsin, Madison, WI 53706, USA}
\affiliation{Institute of Physics, University of Mainz, Staudinger Weg 7, D-55099 Mainz, Germany}
\affiliation{Universit\'e de Mons, 7000 Mons, Belgium}
\affiliation{Bartol Research Institute and Dept. of Physics and Astronomy, University of Delaware, Newark, DE 19716, USA}
\affiliation{Dept.~of Physics, University of Oxford, 1 Keble Road, Oxford OX1 3NP, UK}
\affiliation{Dept.~of Physics, University of Wisconsin, River Falls, WI 54022, USA}
\affiliation{Oskar Klein Centre and Dept.~of Physics, Stockholm University, SE-10691 Stockholm, Sweden}
\affiliation{Dept.~of Astronomy and Astrophysics, Pennsylvania State University, University Park, PA 16802, USA}
\affiliation{Dept.~of Physics, Pennsylvania State University, University Park, PA 16802, USA}
\affiliation{Dept.~of Physics and Astronomy, Uppsala University, Box 516, S-75120 Uppsala, Sweden}
\affiliation{Dept.~of Physics, University of Wuppertal, D-42119 Wuppertal, Germany}
\affiliation{DESY, D-15735 Zeuthen, Germany}

\author{R.~Abbasi}
\affiliation{Dept.~of Physics, University of Wisconsin, Madison, WI 53706, USA}
\author{Y.~Abdou}
\affiliation{Dept.~of Subatomic and Radiation Physics, University of Gent, B-9000 Gent, Belgium}
\author{T.~Abu-Zayyad}
\affiliation{Dept.~of Physics, University of Wisconsin, River Falls, WI 54022, USA}
\author{J.~Adams}
\affiliation{Dept.~of Physics and Astronomy, University of Canterbury, Private Bag 4800, Christchurch, New Zealand}
\author{J.~A.~Aguilar}
\affiliation{Dept.~of Physics, University of Wisconsin, Madison, WI 53706, USA}
\author{M.~Ahlers}
\affiliation{Dept.~of Physics, University of Oxford, 1 Keble Road, Oxford OX1 3NP, UK}
\author{K.~Andeen}
\affiliation{Dept.~of Physics, University of Wisconsin, Madison, WI 53706, USA}
\author{J.~Auffenberg}
\affiliation{Dept.~of Physics, University of Wuppertal, D-42119 Wuppertal, Germany}
\author{X.~Bai}
\affiliation{Bartol Research Institute and Dept. of Physics and Astronomy, University of Delaware, Newark, DE 19716, USA}
\author{M.~Baker}
\affiliation{Dept.~of Physics, University of Wisconsin, Madison, WI 53706, USA}
% ...

\collaboration{IceCube Collaboration}
\noaffiliation

\begin{abstract}
IceCube has become the first neutrino telescope with a sensitivity below the TeV neutrino flux predicted from gamma-ray bursts if GRBs are responsible for the observed cosmic-ray flux above $10^{18}$ eV.
Two separate analyses using the half-complete IceCube detector, one a dedicated search for neutrinos from $p \gamma$-interactions in the prompt phase of the GRB fireball, and the other a generic search for any neutrino emission from these sources over a wide range of energies and emission times, produced no evidence for neutrino emission, excluding prevailing models at 90\% confidence.
\end{abstract}


\maketitle
%\section{Introduction}
Gamma-ray Bursts (GRBs) have long been proposed \cite{waxman95} as one of the most plausible sources of the highest energy cosmic rays, as the observed flux can be entirely explained if the primary engine of the bursts accelerates protons and electrons with comparable efficiencies. The electrons would produce the observed gamma-ray emission by synchrotron emission and, possibly, inverse Compton scattering, while the protons escape to form the high-energy cosmic rays observed at Earth.
Waxman and Bahcall observed\cite{wb97} that, in this case, a potentially detectable flux of high-energy neutrinos is produced by $p \gamma$ interactions when protons and photons coexist in the primary fireball.
The detailed flux predictions are dependent on the fireball parameters; here we use the model by Guetta et al.\cite{guetta2004} to compute these parameters from observations by gamma-ray telescopes.
Past searches with IceCube and other neutrino telescopes have met with negative results \cite{ic22_grb,amanda_grb,super_k_2009}, but have never before had sensitivities at the level of the expected flux.
We search in this work for neutrinos in coincidence with 117 GRBs with half of the IceCube detector complete, and for the first time reach a sensitivity that would yield a positive result given expected fireball parameters, with a 4$\sigma$ expected excess.

%\section{IceCube}
IceCube is a TeV-scale neutrino telescope currently under construction at the South Pole which detects neutrinos by measuring the Cerenkov light from secondary charged particles produced in neutrino-nucleon interactions. 
A total of 5160 Digital Optical Modules\cite{icecube_daq} containing 10-inch photomultipliers and arranged in 86 strings frozen in the ice will make up the full detector; the results presented here were obtained using the first 40 of these strings.
Although capable of detecting multiple flavors of neutrinos from the entire sky, for point sources the detector is sensitive primarily to up-going muons produced in muon neutrino charged-current interactions.
Searches in the muon channel benefit from good angular resolution ($\sim0.7^\circ$ for $E_\nu \agt 10$ TeV) and from the long range of high energy muons (several km at TeV energies), which substantially increases the effective volume of the detector.
By using up-going tracks, the Earth is used to shield against the much larger flux of down-going muons from cosmic ray interactions in the atmosphere. 
Backgrounds from cosmic ray-produced muons and atmospheric neutrinos can be further reduced using the muon energy, as neutrinos from GRBs are expected to have higher energies than from either atmospheric source.

%\section{ Event Reconstruction }

The origin of observed events in IceCube is determined by fitting a track to the hit pattern of the detected Cerenkov light using a maximum likelihood method\cite{muon_reconstruction}. 
The location of the maximum is used as the source of the associated neutrino (collinear with the muon), and the statistical uncertainty in the fit provides an estimate of the uncertainty on the reconstructed direction\cite{paraboloid}.

Due to the stochastic nature of muon energy-loss processes and the rarity of events fully contained within the detector, it is not possible to measure the energy of either the muon or the primary neutrino directly. It is, however, possible to measure the mean energy-loss rate of muons in the detector, which is correlated at high energies with the muon energy and with the original neutrino energy\cite{photorec_icrc}.
The uncertainty of the muon energy using this method is on the order 0.3-0.4 in $\log_{10}E$.

%\section{Burst Sample}

IceCube operated in a 40-string configuration from April 5, 2008 until May 20, 2009.
During that time 129 GRBs were reported in the northern hemisphere via the GRB Coordinates Network\cite{GCN} (GCN).
We assembled a catalog using data from GCN notices and circulars, where the position of the burst was taken from the notice with the lowest reported positional error. For bursts which were localized only by the Fermi Gamma-ray Burst Monitor (GBM), the position was instead taken from the GBM Burst Catalog\cite{gbm_catalog}.
The start and stop times of the prompt gamma-ray phase, $T_{\mathrm{start}}$ and $T_{\mathrm{stop}}$ respectively, were determined by taking the earliest and latest times any satellite reported detecting gamma-rays.
The fluence and spectral information were taken preferentially from Fermi GBM, Konus-Wind, Suzaku WAM, then \emph{Swift}.

Fermi GRBs for which no fluence was reported because the burst was too weak were removed.
GRB080521, GRB081113 and GRB090515 occurred during detector downtime and were removed from the catalog.
GRB090422 and GRB090423 occurred during a preliminary run with 59 strings in operation and will be analyzed later.
The final catalog contained 117 bursts.

\begin{figure}[h]
  \centering
  \includegraphics[width=\linewidth]{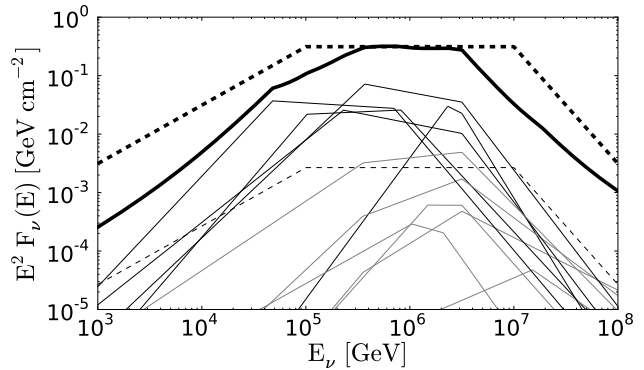}
  \caption{
    The neutrino spectra, including oscillations, of the five brightest GRBs are shown along with eight randomly selected bursts (thin lines).
    A single burst with Waxman 2003 parameters\cite{Waxman2003}, assuming a cosmic ray energy density of $10^{44}$ erg Mpc$^{-3}$ yr$^{-1}$, is shown by a thin dashed line.
    The sum of all 117 individual bursts is shown as a thick solid line along with the Waxman 2003\cite{Waxman2003} prediction in a thick dashed line.
    }    
  \label{fig:neutrino_spectra}
\end{figure}

Neutrino spectra were calculated \cite{guetta2004,ic22_grb} using data from the gamma-ray spectra of individual bursts, or average parameters if no spectral measurements were available.
Definitions of parameters and equations used to calculate neutrino fluence are identical to Appendix A of \cite{ic22_grb}.
Spectra were calculated as power laws with two breaks: a low energy break associated with the break in the photon spectrum, and a high energy break from synchrotron losses of muons and pions (Fig. \ref{fig:neutrino_spectra}).

From the length of gamma emission and energy spectrum, bursts are classified by GCN into two groups (long-soft and short-hard), which may have different underlying sources.
If a burst was not explicitly identified as one class in a GCN notice, we used average values for a short-hard burst if 90\% of the gamma emission was in less than 2 seconds\cite{ic22_grb}, and a long-soft burst otherwise.
Parameters for average long-soft bursts are from \cite{ic22_grb}. For short-hard bursts, we used $L_\gamma^{\mathrm{iso}}= 10^{51}$ erg/s, $t_{\mathrm{var}}= 0.001$ s, and for redshift ($z$) the average of all \emph{Swift} short burst measurements.

%\section{Analysis Overview}
Two independent searches were conducted: one searching for neutrinos with the specific energy spectrum predicted by Guetta et al.\cite{guetta2004} during the period of maximum gamma emission, and the other searching generically for high-energy neutrinos within up to 24 hours of the observed bursts.

%\section{Model-Dependent Analysis}
    
The first of the two analyses, the model-dependent analysis, was designed specifically to find neutrinos produced in $p \gamma$ interactions during the prompt phase of the GRB. Events observed in the detector were reduced by a series of cuts designed to select neutrino-like events, resulting in a data sample of primarily atmospheric neutrinos, an irreducible background for this analysis. We then conducted an unbinned maximum likelihood search\cite{ic22_grb} in which each event passing these cuts was assigned likelihoods of being a signal event (from a GRB) and of being a background event.
Both the signal and background likelihoods for each event $i$ were the product of three independent probability density functions (PDFs) based on direction, arrival time, and muon energy.

The spatial signal PDF was a 2-dimensional Gaussian:
\begin{equation}
\label{eq:pdf_space}
  P^S(\vec{x_i})=\frac{1}{2 \pi (\sigma_{\mathrm{GRB}}^2+\sigma_i^2)}\exp\left(-\frac{|\vec{x}_{\mathrm{GRB}}-\vec{x}_i|^2}{2(\sigma_{\mathrm{GRB}}^2+\sigma_i^2)}\right)
\end{equation}
where $|\vec{x}_{\mathrm{GRB}}-\vec{x}_i|$ is the angle between the reconstructed neutrino direction and the best location of the gamma ray burst provided by GCN, and $\sigma_{\mathrm{GRB}}$ and $\sigma_i$ are the localization uncertainty of the GRB and the muon reconstruction respectively.
The spatial background PDF was computed using a smoothed histogram of all off-source data in detector coordinates, accounting for zenith and azimuth asymmetry in the detector.

The temporal signal PDFs were constant during the prompt phase of the Gamma-Ray Burst (between $T_{\mathrm{start}}$ and $T_{\mathrm{stop}}$), with Gaussian tails of width $T_{\mathrm{stop}} - T_{\mathrm{start}}$ (constrained to at minimum 2 seconds and at maximum 30). The background PDFs were constant in time.

The signal energy PDF was computed from the reconstructed muon energy-loss ($dE/dx$) for neutrinos simulated with the average of the individual burst spectra (Fig. \ref{fig:neutrino_spectra}), while the background energy PDF was computed from the $dE/dx$ distribution of off-source data.

From these likelihoods, we then computed the maximally likely number of signal events. The resulting likelihood ratio (the \emph{test statistic}) was then compared to the distribution from scrambled background datasets to compute the significance of a result.

%\section{Model-Independent Analysis}

As well as looking for neutrinos with properties modeled from measured burst parameters, we conducted an additional search (the model-independent analysis) using wider time search windows and looser event selection criteria, allowing observation of events with late or early arrival times or with unexpected energies due to unanticipated emission mechanisms.

Starting at the interval from -10 to +10 seconds from the GRB trigger time, we expanded a search time window in one second increments in both directions out to $\pm$ one day, looking for a significant excess of neutrinos at each iteration. High correlation between adjacent time windows reduces the trials correction to the significance of any excess to only a few hundred.

Event selection for the model-independent search was based entirely on rejecting misreconstructed downgoing atmospheric cosmic ray muons, which are the dominant background to this analysis, constituting more than 99.9\% of the final 161 million event sample. To avoid assuming a signal neutrino spectrum, no attempt was made to reject the small low-energy background from atmospheric neutrinos.

To ensure that no events were missed due to incorrect assumptions, this analysis was designed to maximize the number of signal neutrinos in the final analysis instead of the significance of an excess. Instead of being selected by hard cuts, events were weighted by their probabilities of being signal neutrinos\cite{MoralesMilagro}. Each probability was the product of the event's point-spread probability density function (Eq. \ref{eq:pdf_space}) and the probability that the event was a neutrino, determined by dividing smoothed histograms of detector data and neutrino simulation in several variables related to reconstruction accuracy. These were then summed in each time window to form the expectation of the on-source signal neutrino density, which was then compared to the expected background value obtained by scrambling the observed data in time.

%\section{Systematic Uncertainties}
Although the use of scrambled data for the background reduces many possible uncertainties, the use of simulation for signal introduces some systematic errors.
The dominant sources of uncertainties in the final limits from both analyses are photon propagation in the ice, the quantum efficiency of the PMTs, and theoretical uncertainties in both the neutrino-nucleon cross section and cross-sections for muon energy-loss processes at high energies.
Depending on the analysis and time interval, the cumulative effect of these uncertainties amounts to 2-13\% and has been included in the final limits using a Bayesian marginalization procedure\cite{FC_systematics}.

%\section{Results}

%\subsection{Model-Dependent Search}

\begin{figure}[ht]
  \centering
  \includegraphics[width=\linewidth]{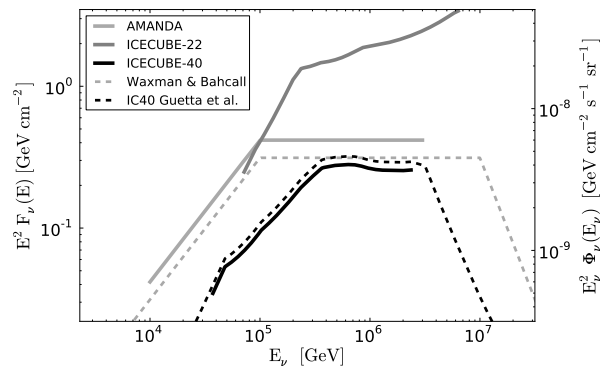}
  \caption{ 
    90\% CL Neyman\cite{pdg2010} upper limit (including systematics) set by model-dependent analysis in solid black with the expected Guetta et al. flux in dotted black.
    22-string IceCube limit\cite{ic22_grb} is in dark gray and AMANDA\cite{amanda_grb} in light gray.
    The Waxman 2003 flux\cite{Waxman2003} is shown for comparison in dotted light gray.
    Diffuse fluxes were obtained from fluences assuming a total of 667 uniformly distributed bursts per year. Fluences are aggregate for 117 bursts.
  }
  \label{fig:mdresults}
\end{figure}

No events were observed in the model-dependent search with a signal to background likelihood ratio greater than one, with 2.99 signal events expected on a background of 0.097. The closest event to its associated GRB was $26^\circ$ from GRB090301A.
This sets a 90\% upper limit of 82\% of the expected flux in the region $37 - 2400$ TeV where 90\% of the events were expected, including a systematic uncertainty of $\sim2\%$ (Fig. \ref{fig:mdresults}).

%\subsection{Model-Independent Search}

\begin{figure}[ht]
  \centering
  \subfigure{\includegraphics[width=\linewidth]{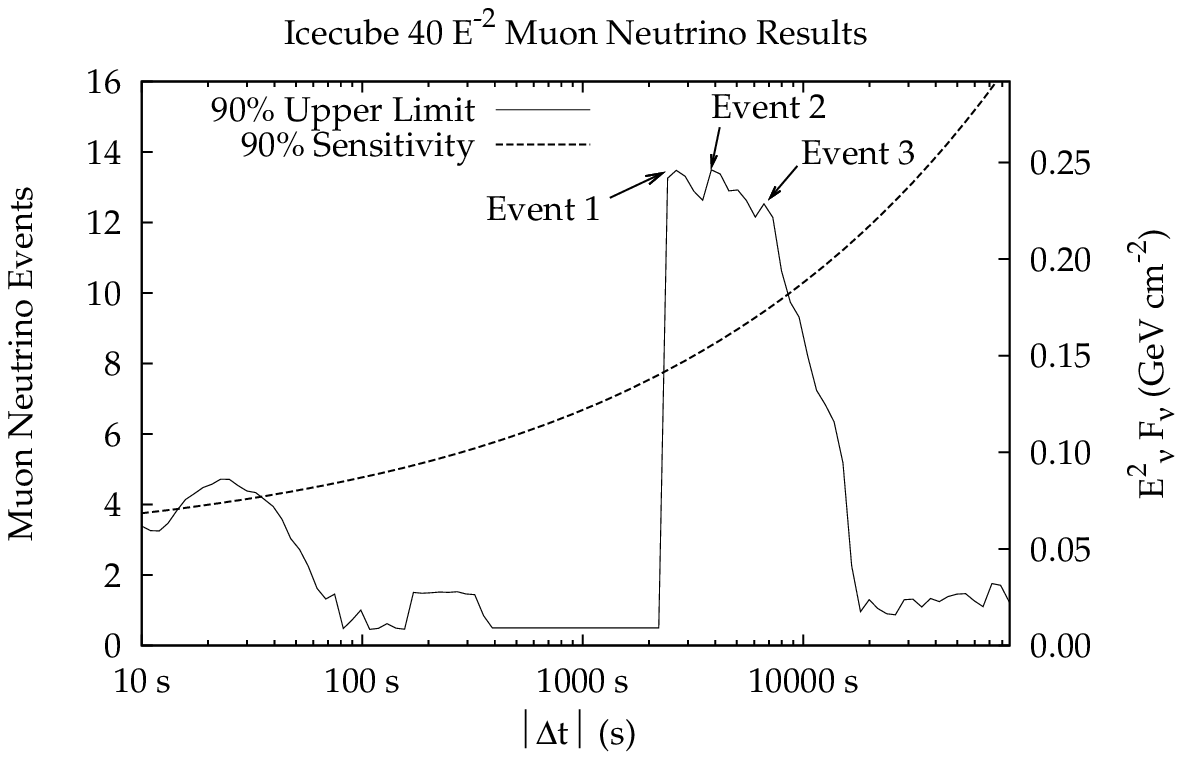}}
  \subfigure{\includegraphics[width=\linewidth]{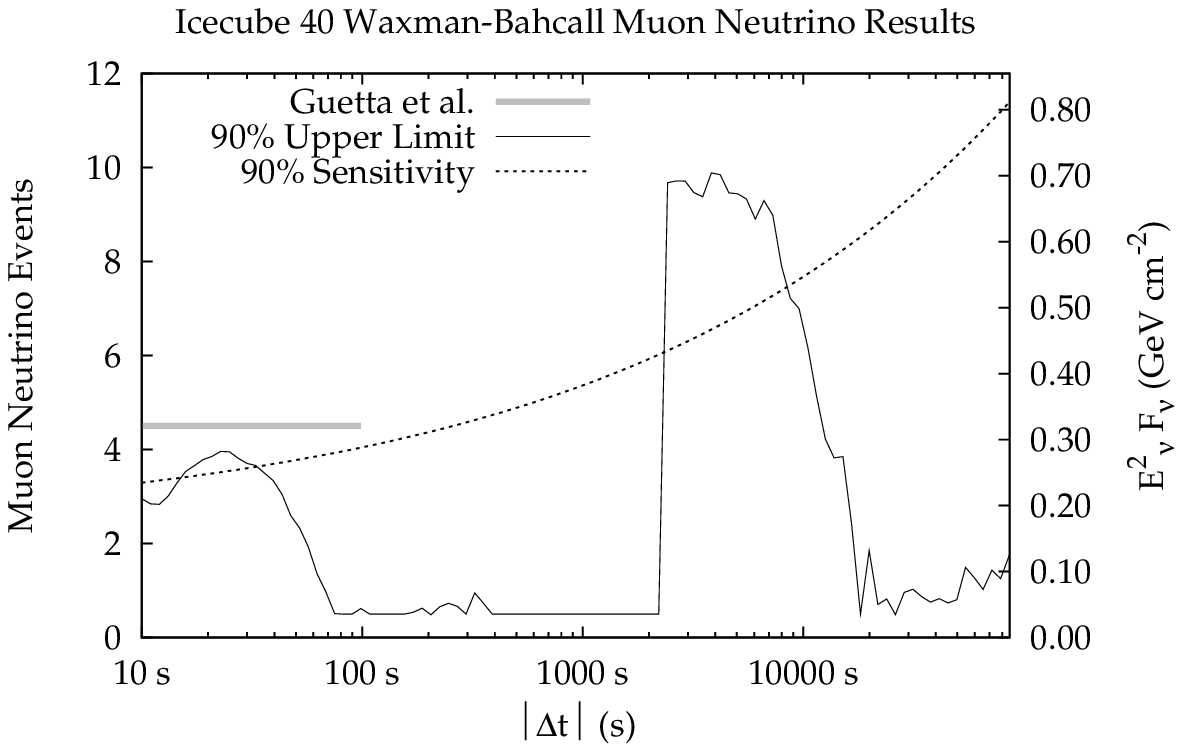}}
  \caption{ 
    90\% CL Feldman-Cousins\cite{FeldmanCousins1998} upper limit (fluence normalization at 642 TeV, the first peak of the expected spectrum) set by the model-independent analysis in each time window for an $E^{-2}$ and for the Guetta et al. spectrum. Systematic errors on the number of events are included. There is an additional $^{+6}_{-2}\%$ uncertainty on the effective area and thus on the right-hand axis. The three sharp peaks between 2000 and 7000 seconds are caused by three low-energy neutrino events consistent with the atmospheric background.
  }
  \label{fig:miresults}
\end{figure}

In the model-independent search, no candidate events were observed in the interval $\pm 2248$ seconds with 4.2 expected from the Guetta et al. calculation. The variation of the upper limit (Fig. \ref{fig:miresults}) with $\Delta t$ reflects statistical fluctuations in the background, as well as the presence of individual events of varying quality. The three most significant of these occurred at $-2249$, $-3594$, and $-6430$ seconds respectively, and were low energy ($\sim1$ TeV) neutrinos consistent with the atmospheric neutrino background. In addition to a constant $^{+6}_{-2}\%$ uncertainty on the effective area (the ratio of fluence to the expected number of events), there is a systematic uncertainty in the limit on the number of expected events that increases with the size of the time window from 0-10\% (included in Fig. \ref{fig:miresults}). This arises from the increased effect of systematic uncertainties in the event selection as the amount of background in the search window increases and the ability to distinguish GRB neutrinos from background events becomes correspondingly more important.

%\section{Discussion}

While the specific neutrino-flux predictions of the fireball model provided by Waxman and Bahcall\cite{wb97} and by Guetta et al.\cite{guetta2004} are excluded (90\% confidence) by this work, we have not yet ruled out the general picture of fireball phenomenology. The neutrino flux we compute for GRBs is determined by the flux of protons accelerated in the fireball, and by the fraction of proton energy transferred to charged pions ($f_{\pi}$). The proton flux can be chosen either such that the energy in gammas and protons is equal or set to the flux of cosmic rays above $10^{18}$ eV, with similar results. $f_{\pi}$ is determined largely by assuming protons are accelerated, in conjunction with the observed low optical thickness of the source. Due to uncertainties in the bulk boost factor and internal structure of the shocks, $f_{\pi}$ may range from 10 - 30\%\cite{GuettaSpadaWaxman2001}, causing an uncertainty of about a factor of 2 on our calculation of the flux, which used $f_{\pi} \approx 0.2$. Future observations by IceCube will push our sensitivity below the level of this theoretical uncertainty on $f_{\pi}$ and allow direct constraints on acceleration of protons to ultra-high energies in Gamma Ray Bursts.

\begin{acknowledgments}
We acknowledge support from the following agencies:
US NSF - Office of Polar Programs,
US NSF - Physics Division,
U. of Wisconsin Alumni Research Foundation,
the GLOW and OSG grids;
US DOE, NERSCC, the LONI grid;
NSERC, Canada;
Swedish Research Council,
Swedish Polar Research Secretariat,
SNIC,
K. and A. Wallenberg Foundation, Sweden;
German Ministry for Education and Research,
Deutsche Forschungsgemeinschaft;
%Research Department of Plasmas with Complex Interactions (Bochum), Germany;
FSR,
FWO Odysseus,
IWT,
BELSPO, Belgium;
%University of Oxford, United Kingdom;
Marsden Fund, New Zealand;
JSPS, Japan;
SNSF, Switzerland;
A.~Gro{\ss} is supported by the EU Marie Curie OIF Program, 
J.~P.~Rodrigues by the Capes Foundation, Brazil,
N.~Whitehorn by the NSF GRFP.
\end{acknowledgments}

\end{document}